# Integrated Transmission and Distribution Expansion Planning Considering Customer Actions and Distributed Energy Resources

Abhinav Ayri[1], Haotian Yao[2], Mostafa Farrokhabadi[2], Hamidreza Zareipour[2]

[1]*FortisAlberta, Canada* [2]*University of Calgary, Canada*

ab.ayri@fortisalberta.com, haotian.yao@ucalgary.ca, mostafa.farrokhabadi@ucalgary.ca, hzareipo@ucalgary.ca

*Preprint. This is the author's version of a paper accepted for presentation at the CIGRE Energy Forum, Calgary, Alberta, Canada, 21–24 September 2026 (Paper 10200, Study Committee C2, Preferential Subject 5 – Distribution Management).*

***Abstract*—**The continued integration of distributed energy resources (DERs) motivates research into integrated system planning. While existing literature shows that DERs can reduce system costs, it often overlooks customer behaviour, particularly DER adoption driven by cost savings and incentives. System cost allocation can influence these decisions, affecting DER uptake, grid injections, and overall system efficiency. Therefore, integrated system planning should consider both system- and customer-level benefits of DERs. This paper proposes a multi-step framework that combines an integrated planning model with cost allocation and retailer business models. The integrated planning model minimizes total system costs subject to network constraints, while the cost allocation and retailer models capture customer responses to costs and DER opportunities. By coordinating these models, the framework represents how customer behaviour influences planning outcomes. The proposed approach is demonstrated on a 36-bus system, with results showing that incorporating customer decision-making reduces overall system costs, lowers reliance on transmission-connected generation, and supports more realistic system planning.

## I. INTRODUCTION

Transmission and distribution systems have traditionally been planned independently, as power flows were predominantly unidirectional from transmission to distribution [1]. However, the increasing penetration of distributed energy resources (DERs), such as rooftop photovoltaic systems and battery storage, has introduced bidirectional power flows and significantly altered transmission-distribution interactions [2]. Integrated transmission and distribution planning (integrated system planning) has been shown to capture DER benefits and reduce overall system costs [1]-[2].

Existing literature on integrated system planning primarily focuses on minimizing investment and operational costs under uncertainty, including renewable integration [3]-[7]. While DERs are often included, they are typically modeled as controllable resources from a system perspective, and their adoption and operation are rarely treated as outcomes of customer decision-making.

In practice, DER investment and operation are driven by customer-level factors, including electricity prices, tariff structures, and perceived financial benefits [8]. Additionally, planners often have limited visibility and control over these resources and may continue proposing traditional system upgrades.

Despite extensive evidence that DERs can reduce system costs, existing planning models rarely capture customer-driven adoption and operation. This paper proposes a multi-stage framework that integrates planning, cost allocation, and customer decision models to better represent customer-driven DER behaviour and its influence on system planning outcomes.

## II. THE PROPOSED PLANNING MODEL

This paper presents the transmission-distribution expansion planning framework proposed in [11], comprising planning, cost allocation, and customer decision stages, based on the models in [3], [9], and [10], respectively. A key feature is the explicit modeling of customer decisions to adopt and operate DERs based on energy needs and incentives. Customer actions are incorporated into the integrated planning model as updated hourly load profiles in a subsequent stage.

### *A. First planning stage*

The first step is to run the integrated planning model to establish a baseline without DERs. As described in [11], the model combines transmission and distribution expansion planning to capture system coordination. The objective is to minimize total annualized investment and operating costs:

$$c^T = c^{I,T} + c^{O,T} + c^{I,D} + c^{O,D} \tag{1}$$

subject to transmission and distribution system constraints. The objective captures transmission and distribution investment ($c^{I,T}$, $c^{I,D}$) and operating costs ($c^{O,T}$, $c^{O,D}$). Investment costs include installing transmission-connected generation and candidate transmission and distribution lines, while operating costs reflect production costs of existing transmission-connected generation and load-shedding of demand. DER adoption and operation are determined in the customer decision stage and subsequently incorporated into the planning model as updated load profiles.

The model is subject to transmission and distribution constraints, including power balance equations for both systems. The transmission system is represented using DC power flow, while the distribution system is represented using the Linearized DistFlow formulation [12]. Additional constraints include flow limits for existing and candidate transmission and distribution lines, existing and candidate generation capacity limits, bounds on load-shedding, and voltage magnitude limits for distribution nodes. The full formulation is provided in [11].

### *B. Cost allocation stage*

The active power and load-shedding values obtained from the integrated planning model in the first planning stage are used to determine the cost and price per node of each distribution customer consuming electricity from the system. These values are used to identify the customers most likely to adopt DERs and to inform the determination of DER capacity injected or absorbed in the customer decision stage. The model assigns costs to customers based on their demand and the power supplied by the system to meet this demand.

Distribution customer costs reflect both transmission and distribution components (i.e., integrated costs). The cost allocation model outputs the integrated cost and price at each distribution node:

$$\text{Integrated cost}_n = \text{Transmission cost}_n + \text{Distribution cost}_n \tag{2}$$

$$\text{Integrated price}_n = \frac{\text{Integrated cost}_n}{\text{Net load demand}_n} \tag{3}$$

In (2), the transmission cost reflects the power supplied from the transmission system to distribution node $n$, while the distribution cost reflects the power flows within the distribution system required to serve that demand. The integrated price is calculated as the ratio of the integrated cost to net load demand. Customers with the five highest costs and prices are identified as candidates for DER adoption. The full formulation of this stage is provided in [11].

### *C. Customer decision stage*

Customer DER adoption is modeled using a retailer-based economic framework [10]-[11]. Inputs include hourly load data from the first planning stage and integrated prices from the cost allocation stage. Outputs include retailer payments to customers, retailer revenues and profits, payments to distribution companies, and DER adoption across multiple decision intervals. The adopted DER capacity is determined as:

$$size^{*}_{der} = \begin{cases} size^{k}_{der} & \text{if } lcoe_k \leq lcoe_{BaU} \\ 0 & \text{otherwise} \end{cases} \tag{4}$$

$$lcoe_k = \min\left(lcoe_{off}, lcoe_{BaU}, lcoe^{i}_{der}\right) \tag{5}$$

DER adoption is modeled by comparing the levelized cost of energy (LCOE) of different photovoltaic (PV) and battery configurations ($lcoe^{i}_{der}$) with the LCOE of not adopting new DERs ($lcoe_{BaU}$). The equations also account for the possibility of customers defecting from the grid ($lcoe_{off}$). The LCOE includes PV and battery capital costs, operations and maintenance (O&M) costs, cumulative prior investments, and electricity bills. Customers are also constrained to generate sufficient energy to meet their annual demand [10]. Customer operation is modeled using a household energy management system to determine hourly net power exchange (either consumption from or injection into the grid) and to update load profiles for the planning model. These outputs are subsequently used as inputs to the planning model in a later stage:

$$P^{h,t}_{net} = P^{h,t}_{d} - P^{h,t}_{g} + P^{h,t}_{b} \tag{6}$$

where $P^{h,t}_{net}$, $P^{h,t}_{d}$, $P^{h,t}_{g}$, and $P^{h,t}_{b}$ represent net power exchange, demand, PV generation, and battery charge/discharge, respectively.

### *D. Second planning stage*

Using the updated load profile from the customer decision stage, the integrated planning model described in the first planning stage is re-run to compare scenarios with and without customer-driven DER adoption. The framework ultimately compares a planner-driven scenario that excludes DER integration with one that reflects customer-driven adoption and operation.

## III. RESULTS

The framework was tested on a 36-bus system consisting of 6 transmission nodes and 30 distribution nodes to evaluate the impact of customer actions on system costs. Figure 1 shows the abridged network, in which each distribution system comprises 15 nodes [11]. Nodes in Distribution System 1 are designated as 4-1, 4-2, ..., 4-15, while nodes in Distribution System 2 are designated as 5-1, 5-2, ..., 5-15.

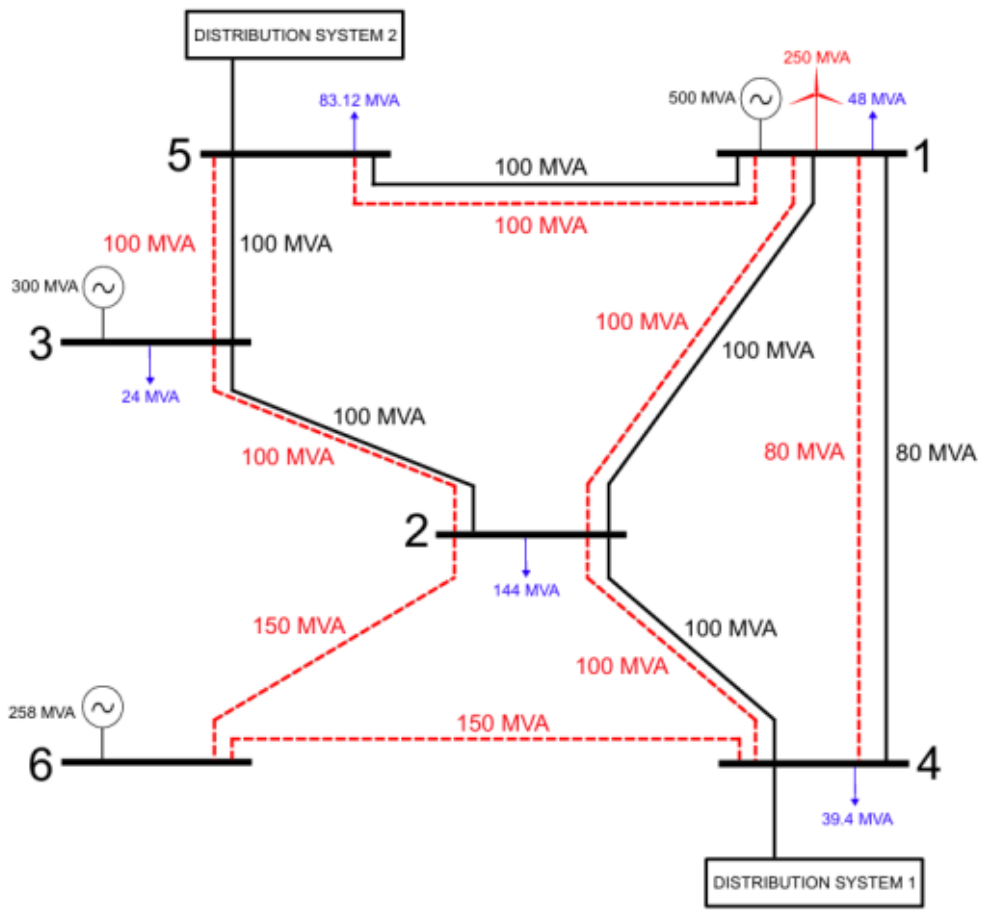

***Figure 1. 36-node system.***

Data for the 36-bus test system was adopted from [3], including peak demand, line impedances, generation investment and production costs, line investment costs, and load-shedding costs. Hourly load data was obtained from the Alberta Electric System Operator (AESO) website from May 1, 2020, to May 1, 2021 [13]. PV, battery, and O&M costs were taken from the National Renewable Energy Laboratory (NREL) Annual Technology Baseline [14] for the period 2022-2050. These costs decline annually (under either conservative or advanced assumptions) and are used in the retailer business model to determine DER adoption across decision intervals. Further details are provided in [11].

### A. First planning, cost allocation, and customer decision stage results

Based on the first planning and cost allocation stages, nodes 4-1, 4-3, 4-7, 4-14, and 5-7 were identified as high-cost customers and selected for DER adoption. Four scenarios were investigated for DER adoption by distribution customers. Scenario 1 considered customers adopting DERs with conservative PV, battery, and O&M cost estimates with no feed in tariff incentive, Scenario 2 considered adoption with advanced cost estimates and no feed in tariff incentive, Scenario 3 considered adoption with conservative cost estimates and a feed-in tariff incentive of \$0.3/MWh, and Scenario 4 considered adoption with advanced cost estimates and a feed-in tariff of \$0.3/MWh.

Figure 2 summarizes the adopted DER capacities. Adoption increases from Scenarios 1 to 4, with feed-in tariffs (Scenarios 3 and 4) yielding the highest uptake. Specifically, DER adoption increases by approximately 75% from Scenario 1 to Scenario 2, and by approximately 300% from Scenario 1 to Scenarios 3 and 4. Uniform adoption trends arise because all customers face identical cost assumptions and incentives. Scenarios 3 and 4 yield identical capacities, as PV adoption is capped by annual consumption, and advanced cost estimates therefore do not result in additional adoption. In all scenarios, customers do not adopt battery storage.

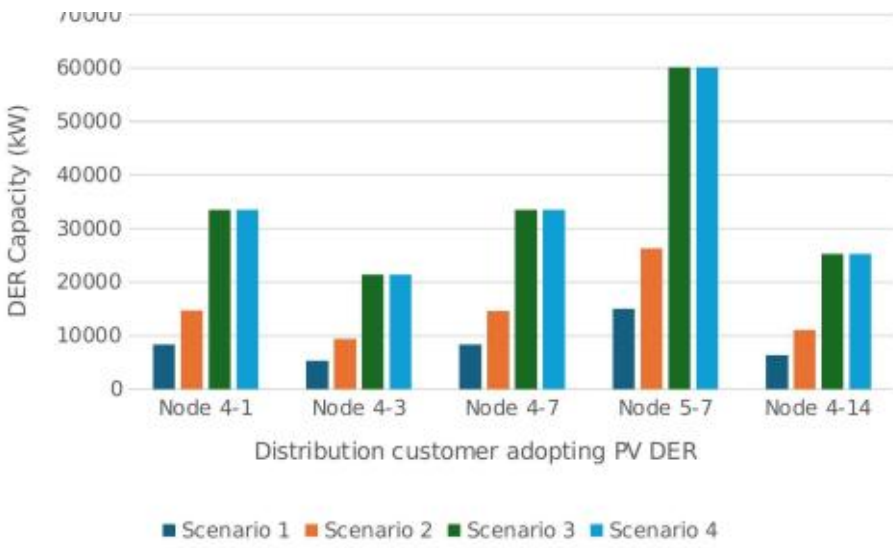

*Figure 2. PV DER adoption results across scenarios.*

### B. Second planning stage and additional results

After the customer decision stage, the second planning stage was performed. Updated hourly load profiles from Scenarios 1 to 4 were incorporated into the original load data. Figure 3 summarizes system costs across all scenarios.

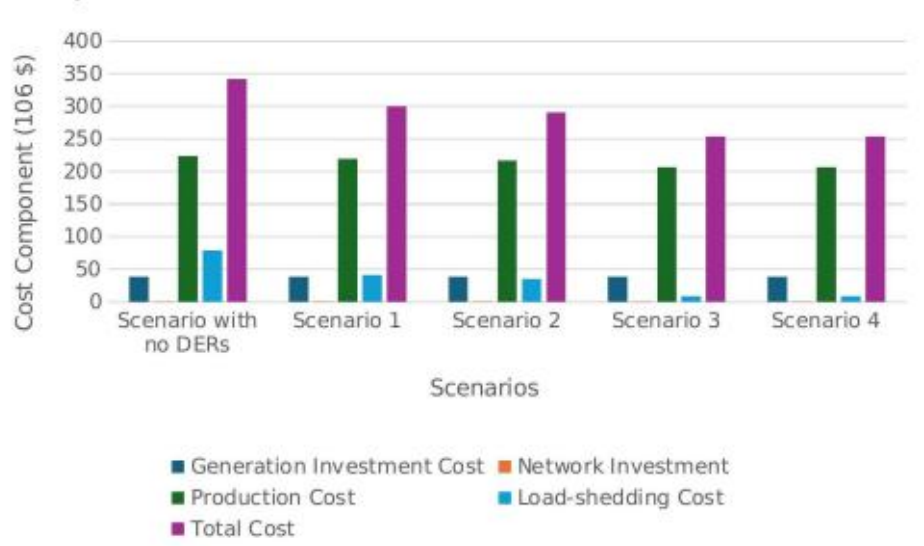

*Figure 3. Integrated cost results across scenarios.*

In the no DERs scenario, production and load-shedding costs dominate, indicating reliance on transmission-connected generation and limited capacity to meet demand. Across Scenarios 1 to 4, these costs decrease, with the largest reductions occurring in Scenarios 3 and 4. In these scenarios, production and load-shedding costs decrease by 7.61% and 89.4%, respectively, while overall system costs decline by approximately 25.7%. Increased DER adoption reduces reliance on transmission supply and lowers system costs. The reduction in load-shedding costs further indicates that DERs help meet distribution-level demand.

Results from the second planning stage also provide active power values across the model time horizon. These values are compared at a given instance between the scenario with no DERs and Scenario 4 to assess the impact of DERs on system power flows. This comparison is shown in Figure 4 [11].

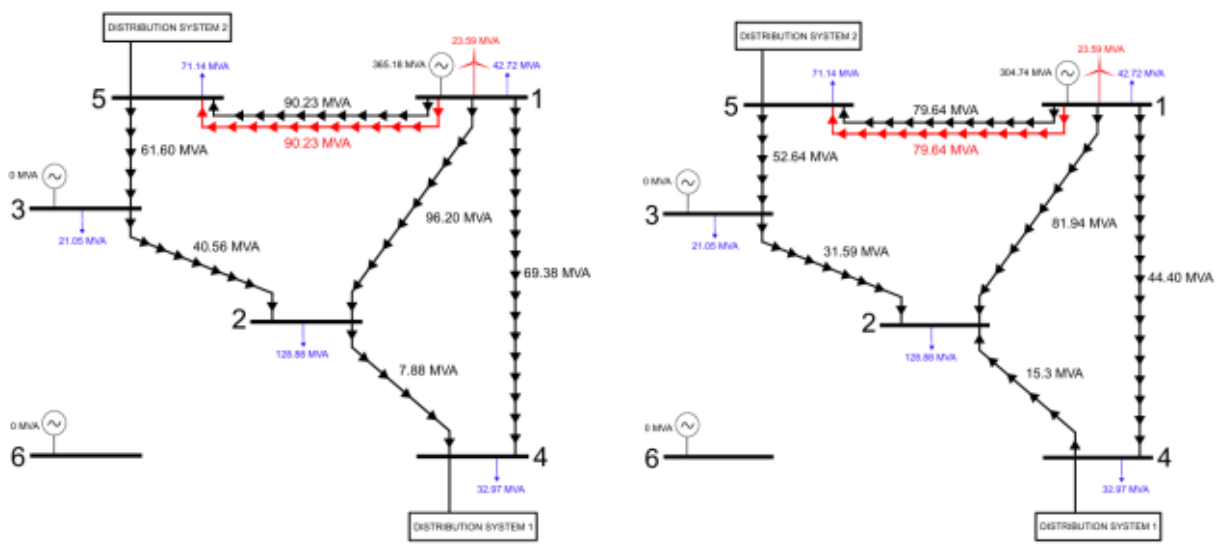

*Figure 4. Comparison of active power flows for scenario with no DERs (left) with Scenario 4 (right).*

For this instance, loading decreased across most of the transmission lines. The total apparent power flow into node 4 (serving Distribution System 1) dropped from 77.24 MVA (69.36 MVA + 7.88 MVA) to 29.1 MVA (44.40 MVA – 15.3 MVA), corresponding to a 62.3% reduction. Similarly, the total flow into node 5 (serving Distribution System 2) decreased from 118.86 MVA (90.23 MVA + 90.23 MVA – 61.60 MVA) to 106.64 MVA (79.64 MVA + 79.64 MVA – 52.64 MVA), representing a 10.3% reduction. These reductions occur because DERs in Scenario 4 supply power locally, thereby reducing transmission loading.

Additionally, the cost allocation stage was re-run using the results from the second planning stage to evaluate changes in the integrated costs of distribution customers. Figure 5 summarizes the integrated costs of distribution customers in the scenario with no DERs and Scenario 4.

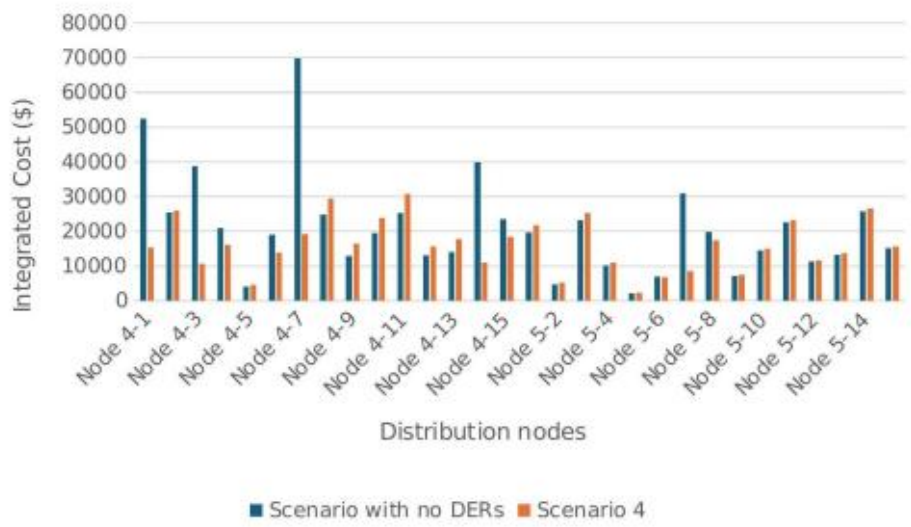

*Figure 5. Integrated cost comparison between scenario with no DERs and Scenario 4.*

The integrated costs for the DER-adopting customers (nodes 4-1, 4-3, 4-7, 4-14, and 5-7) decrease significantly from the scenario with no DERs to Scenario 4. Integrated costs also declined for a subset of non-adopting customers, specifically nodes 4-4, 4-6, 4-15, 5-6, and 5-8, while costs for the remaining non-adopting customers increased. The changes observed in Scenario 4 indicate that customers adopting DERs under a non-zero feed-in tariff select capacity levels and inject power at a value that reduce both their own integrated costs and those of nearby non-adopting customers. These customers are electrically adjacent to the DER-adopting nodes: node 4-4 to node 4-3, node 4-6 to node 4-7, node 4-15 to node 4-14, and nodes 5-6 and 5-8 to 5-7. Thus, a portion of the integrated costs that would otherwise be allocated to DER-adopting and their adjacent non-adopting nodes are redistributed to the other non-adopting nodes.

## IV. CONCLUSION

The main contribution of this paper is a planning framework that explicitly models customer DER adoption and operation based on economic signals. Results show that, within the proposed framework, higher DER uptake improves cost efficiency and reliability, reduces transmission loading, and lowers costs for both DER-adopting and some non-adopting customers. The substantial reduction in load shedding highlights an important reliability benefit, namely that customer DERs help alleviate generation shortfalls during constrained conditions. Although customers are not explicitly compensated for these reliability contributions, the benefits accrue implicitly to the system through reduced costs.

The framework also provides insights for policymakers and regulators. Feed-in tariffs encourage customer participation and increase DER injections, indirectly improving system reliability. However, the framework assumes that customers respond solely to economic incentives, with reliability benefits emerging as a byproduct. This separation between customer behaviour and system planning may not fully reflect operational practices, where reliability remains a core responsibility of system operators.

Future work should explore market and regulatory mechanisms that explicitly compensate customers for reliability services. Improved alignment between customer incentives and system needs could further reduce costs, relieve network loading, and ensure a more balanced allocation of benefits while maintaining planning oversight.

The framework in this paper is also sequential, with system planning first conducted without DERs and customer decisions introduced in a subsequent stage. As a result, there is a disconnect between planning outcomes and customer-driven DER adoption, and the framework does not fully capture the feedback between system decisions and customer behaviour. Consequently, the resulting system costs and DER adoption levels may not reflect a fully coordinated equilibrium. Future work could address this limitation by developing a unified, iterative framework that jointly models planning and customer decision-making. While the framework is demonstrated on a benchmark test system, future work should also apply the methodology to real transmission and distribution networks to evaluate performance under practical planning conditions and to further validate the observed benefits of customer-driven DER adoption.

Overall, the results demonstrate that realizing the full value of DERs requires accounting for both system- and customer-level benefits within integrated planning frameworks.